\documentclass[11pt]{article}

\usepackage{calc}
\makeatletter
\@addtoreset{equation}{section}

\makeatother

\usepackage{amssymb,amsfonts,amsmath,pifont,amscd}

\def\beq{\begin{equation}}
\def\eeq{\end{equation}}
\def\bea{\begin{align}}
\def\eea{\end{align}}
\def\Tr{{\rm Tr}}

\renewcommand{\t}{\theta}

\renewcommand{\d}{\delta}
\newcommand{\D}{\Delta}

\newcommand{\s}{\sigma}
\newcommand{\vph}{\varphi}

\newcommand{\cd}{\mathcal{D}}
\newcommand{\bsp}{\boldsymbol{\varphi}}
\newcommand{\bspsi}{\boldsymbol{\psi}}
\newcommand{\prt}{\partial}
\newcommand{\bsD}{\boldsymbol{\Delta}}

\newcommand{\non}{\nonumber \\}
\newcommand{\bsx}{\boldsymbol{x}}
\newcommand{\psiu}{\underline{\psi}}
\newcommand{\tp}{\tilde{\varphi}}
\newcommand{\tD}{\tilde{D}}
\newcommand{\tpsione}{\tilde{\psi_1}}
\newcommand{\tpsitwo}{\tilde{\psi_2}}

\newcommand{\tf}{\tilde{f}}

\title{Matrix formulation of superspace on 1D lattice with two supercharges}
\author{%
    \authorfont
    Sergio Arianos,
    Alessandro D'Adda,
    Alessandra Feo, \\
    Noboru Kawamoto
    \quad and\quad
    Jun Saito}
\date{May, 2008}

\begin{document}

\renewcommand{\thefootnote}{\fnsymbol{footnote}}
\newcommand{\inst}[1]{\mbox{$^{\text{\textnormal{#1}}}$}}
\begin{flushright}
DFTT 15/2008\\
EPHOU 08-003\\
May, 2008
\end{flushright}
\bigskip\bigskip
\begin{center}
{\LARGE
Matrix formulation of superspace on 1D lattice\\ with two supercharges}\\[8ex]
%
{\large
Sergio Arianos\inst{a}\footnote{\texttt{arianos@to.infn.it}},
Alessandro D'Adda\inst{a}\footnote{\texttt{dadda@to.infn.it}},
Alessandra Feo\inst{a}\footnote{\texttt{feo@to.infn.it}},
Noboru Kawamoto\inst{b}\footnote{\texttt{kawamoto@particle.sci.hokudai.ac.jp}}\\[1ex]
{\normalsize and}
Jun Saito\inst{b}\footnote{\texttt{saito@particle.sci.hokudai.ac.jp}}}\\[4ex]
%
{\large\itshape
\inst{a} INFN sezione di Torino, and\\
Dipartimento di Fisica Teorica,
Universita di Torino\\
I-10125 Torino, Italy\\[3ex]
\inst{b} Department of Physics, Hokkaido University\\
Sapporo, 060-0810 Japan}
\end{center}
\bigskip\bigskip
\setcounter{footnote}{0}
\renewcommand{\thefootnote}{\arabic{footnote}}

\begin{abstract}
Following the approach developed by some of the authors in recent papers and using a matrix representation for the superfields, we formulate an exact supersymmetric theory with two supercharges on a one dimensional lattice.
In the superfield formalism supersymmetry transformations are uniquely defined and do not suffer of the ambiguities recently pointed out by some authors. The action can be written in a unique way and it is invariant
under all supercharges.  A modified Leibniz rule applies when supercharges
act on a superfield product and the corresponding Ward identities take
a modified form but  hold exactly at least at the tree level, while their validity in presence of radiative corrections is still an open problem and is not considered here.
\end{abstract}

\section{Introduction}
A consistent formulation of supersymmetry on the lattice is a long standing
problem (see \cite{Feo:2002yi} and references therein).
Older papers on the subject make use of schemes that break supersymmetry
explicitly on the lattice and recover it in the continuum limit. This however
leads in general to fine tuning problems.
In recent years a number of approaches that allow to preserve exactly one
supersymmetry on the lattice in theories with an extended supersymmetry have
been proposed
\cite{Kaplan:2002wv,Catterall:2001fr,Sugino:2003yb}.
A more ambitious approach, which we follow in the present paper, aiming to
preserve exactly all supersymmetries in some extended supersymmetric model,
was also proposed
\cite{D'Adda:2004jb,D'Adda:2004ia,D'Adda:2005zk,D'Adda:2007ax}%
\footnote{while for the $N=1$ Wess-Zumino model in 4 dimensions an exact
lattice formulation has been achieved using the Ginsparg-Wilson formulation
\cite{Ginsparg:1981bj} which gives exact Ward-Takahashi identities at fixed
lattice spacing \cite{Bonini:2004pm}.
}.
See also related works on the subject
\cite{Dondi:1976tx,Elitzur:1982vh,Nishimura:1997vg,Montvay:2001aj,Onogi:2005cz,Bietenholz:1998qq}
and reviews at lattice conferences
\cite{Kaplan:2003uh,Catterall:2005eh,Giedt:2007hz,Feo:2002yi,Feo:2004kx}.

There are two important well known obstacles in formulating a supersymmetric
lattice theory: Firstly, Poincar\'e invariance is lost on the lattice, thus
supersymmetry is lost as well.
In particular derivatives are replaced on the lattice by finite differences
which do not satisfy the Leibniz rule when applied to a  product of fields
\cite{Dondi:1976tx}.
Secondly, a naive regularization leads to the doubling problem.
This results in the wrong number of degrees of freedom and therefore in a
violation of the balance between fermions and bosons.

The approach we are going to follow makes use of Dirac-K\"ahler twisting procedure
developed in the continuum formulation~\cite{Kawamoto:1999zn}.
Its main feature however is the use of an extended lattice where
the standard links, corresponding to the discrete elementary
translations on the lattice, are implemented by "fermionic" links
that correspond to the action of supersymmetry charges. The need
for these extra links is the result of the modified "shifted"
Leibniz rules, that both translations and supersymmetry
transformation have for consistency to satisfy on a lattice when
acting on a product of (super)fields. The structure of the
extended lattice thus reflects the structure of the twisted
supersymmetry algebra, and a solution is found only for some
specific extended superalgebras which are consistent with the
twisted supersymmetry algebra, like the $N=2$ superalgebra for
$D=2$ \cite{D'Adda:2004jb,D'Adda:2005zk}, the $N=4$ superalgebra
for $D=3$ \cite{D'Adda:2007ax,Nagata:2007mz}, and the $N=4$
superalgebra for $D=4$ \cite{D'Adda:2005zk}. The above mentioned
two obstacles are overcome in this formulation: The twisted
lattice supersymmetry algebra with the difference operator is
realized algebraically by introducing the shifted Leibniz rule.
The extra degrees of freedom of species doublers of the lattice
chiral fermions, which were usually identified as flavor degrees
of freedom, can be identified as extended supersymmetry degrees of
freedom. From the consistency, the supersymmetry charges are in
this approach associated to links, rather than to sites. In
connection with this point a number of criticisms were put forward
\cite{Bruckmann:2006ub,Bruckmann:2006kb}, with the claim that the
link nature of supercharges leads to inconsistencies and
ambiguities in the definition of the supersymmetry
transformations. Very recently \cite{Damgaard:2007be} it was shown
that this approach fits within the scheme of Kaplan's orbifold
formulation \cite{Kaplan:2002wv} and the actual invariance of the
action proposed in \cite{D'Adda:2005zk} under all susy charges has
been questioned.

In order to investigate the above issues we considered a simple
one dimensional supersymmetric model with two supercharges
\cite{Cooper:1982dm}, the
same model already considered by Bruckmann and de Kok in
\cite{Bruckmann:2006ub}.
In this context we develop a matrix representation of superfields on the
lattice, already anticipated in  \cite{Arianos:2007nv}, and show using the
superfield formalism that the lattice action is invariant under all
supersymmetry charges and that supersymmetry
transformations are consistently and unambiguously determined in terms of superfield
on
the lattice, thus overcoming the objections of \cite{Bruckmann:2006ub}.
Since (super)symmetry transformations on the lattice make use of modified
Leibniz rules, supersymmetry cannot be expressed simply as a change of
variables in the functional integral and the existence of Ward identities
showing that it is exactly preserved at the quantum level is not obvious.
We  then decided to investigate tree level Ward identities for the free action
(including
the mass term) and found that exact Ward identities hold in that case showing
that lattice supersymmetry is really a symmetry of the action.
Such Ward identities are "modified" in a way that is reminiscent of the modified
Leibniz rules, and reflect the particular nature of the lattice symmetry.
The question of whether modified Ward identities can be derived also in the case
of interaction, namely at loop level, is still unanswered and is being
investigated.

The paper is organized as follows: In Section 2 we describe the
one dimensional $N=2$ supersymmetric model  in the continuum. In
Section 3 we give some general properties of one dimensional
lattice theories and in particular the modified Leibniz rule
obeyed by finite difference operator when acting on a product of
fields on the lattice. The generalization of this rule to the
supersymmetric case is one of the main ingredients of our
formulation. The superfield formulation of the  $N=2$ one
dimensional supersymmetric model on the lattice, including the
algebra and the supersymmetry transformations, is presented in
Section 4.  In Section 5 we construct the supersymmetric lattice
action in terms of superfields and also in terms of the component
fields. Section 6 is dedicated to the study of the modified Ward
identities and some examples to tree level are given. Conclusions
on Section 7.

\section{The model}
\label{the-model}
The model we are going to discuss in this paper is a one dimensional
supersymmetric model with two supersymmetry charges \cite{Cooper:1982dm},
whose lattice formulation was recently discussed in ~\cite{Bruckmann:2006ub}.
Its supersymmetry algebra is given, in terms of Majorana SUSY charges $Q_1$ and
$Q_2$ by:
\begin{align}
Q_1^2&=Q_2^2= P_x, \nonumber \\
\{Q_1,Q_2\}&=0,~~~~[P_x,Q_1]=[P_x,Q_2]=0,
\label{susyalgebra}
\end{align}
where $P_x$ is the generator of translations in the one-dimensional space-time
coordinate $x$\footnote{The coordinate $x$ is the Wick rotated time coordinate
$t \rightarrow i x$, so that we are actually describing a euclidean formulation.}:
\beq
P_x = \frac{\partial}{\partial x}.
\label{hamilt}
\eeq
A superspace representation of the algebra may be given in terms of two
Grassmann odd, real coordinates $\theta_1$ and $\theta_2$, namely:
\beq
Q_1 = \frac{\partial}{\partial \theta_1} + \theta_1 \frac{\partial}{\partial
x},~~~~~~~~~~Q_2=\frac{\partial}{\partial \theta_2} +
\theta_2 \frac{\partial}{\partial x}.
\label{susycharges}
\eeq
The field content of the theory is described by a hermitian superfield
$\Phi(x,\theta_1,\theta_2)$:
\beq
\Phi(x,\theta_1,\theta_2) = \vph(x) + i \theta_1 \psi_1(x) + i \theta_2
\psi_2(x) + i \theta_2 \theta_1 D(x),
\label{superfield}
\eeq
where $\psi_1$ and $\psi_2$ are Majorana fermions.
The supersymmetry transformations of the superfield $\Phi$ are given by:
\beq
\delta_j \Phi = [ \eta_j Q_j, \Phi ]~~~~~~~~~~j=1,2,
\label{susytrans}
\eeq
where $\eta_i$ are the Grassmann odd parameters of the transformation.
In terms of the component fields eq. (\ref{susytrans}) reads:
\begin{align}
\delta_j \vph &= i \eta_j \psi_j, \\
\delta_j \psi_k&= i \delta_{j,k} \eta_j \partial_x \vph + \epsilon_{jk} \eta_j D,
\\ \delta_j D &= - \epsilon_{jk} \eta_j \partial_x \psi_k.
\label{susycomp}
\end{align}
It is important to note that since the supersymmetry transformations are defined in
(\ref{susytrans}) as commutators, supersymmetry transformations of superfields
products obey ordinary Leibniz rule:
\beq
\delta_i (\Phi_1 \Phi_2) = (\delta_i \Phi_1) \Phi_2 + \Phi_1 (\delta_i \Phi_2).
\label{LBr}
\eeq
In order to write a supersymmetric action we need to introduce the
superderivatives, defined as
\beq
D_j = \frac{\partial}{\partial \theta_j} - \theta_j \frac{\partial}{\partial
x},
\label{susyder}
\eeq
which anticommute with the supersymmetry charges $Q_j$ and satisfy the algebra:
\beq
D_j^2 = - \frac{\partial}{\partial x},~~~~~~~~~~\{D_1,D_2\} = 0.
\label{sderalg}
\eeq
The supersymmetric action can then be defined in terms of the superfield $\Phi$
as:
\beq
\int dx d\theta_1 d\theta_2 \left[ \frac{1}{2} D_2 \Phi D_1 \Phi + i V(\Phi)
\right],
\label{action}
\eeq
where $V(\Phi)$ is a superpotential that we will choose, as in
~\cite{Bruckmann:2006ub}, to be of the form:
\beq
V(\Phi) = \frac{1}{2} m \Phi^2 + \frac{1}{4} g \Phi^4.
\label{potential}
\eeq
By integrating over $\theta_1$ and $\theta_2$ in (\ref{action}) one can obtain
the action written in terms of the component fields, namely:
\begin{align}
S=& \int dx \{ \frac{1}{2} \left[  (\partial_x \vph)^2 - D^2 -
\psi_1 \partial_x \psi_1 - \psi_2 \partial_x \psi_2 \right] \non
-& m(i \psi_1 \psi_2 + D \vph) - g ( 3 i \vph^2 \psi_1 \psi_2 + D \vph^3 ) \}.
\label{actioncomp}
\end{align}
The auxiliary field $D$ appears quadratically in (\ref{actioncomp}) and can
therefore be integrated out, leading to the on-shell action:
\begin{align}
S=& \int dx \{ \frac{1}{2} \left[  (\partial_x \vph)^2 - \psi_1 \partial_x
\psi_1 - \psi_2 \partial_x \psi_2 +m \vph^2 - 2im\psi_1 \psi_2 \right] \non
  +&mg \vph^4 + \frac{g^2}{2} \vph^6 -3ig \vph^2 \psi_1 \psi_2 \}.
\label{actioncomp2}
\end{align}

\section{One dimensional lattice and modified Leibniz rules}

We discuss in this section some general properties of one dimensional lattice
theories, without introducing supersymmetry, and in particular the modified
Leibniz rule obeyed by the finite difference operator when acting on a product
of fields on the lattice. The generalization of this rule to the supersymmetric
case is one of the main ingredients of our formulation.
To begin with, let us introduce a matrix notation for fields on a one
dimensional lattice.

Consider a one dimensional lattice with $N$ sites and periodic
boundary conditions. Let $a$ be the lattice spacing. The $N$ sites of the
lattice will be labeled by a coordinate $\bsx = r a$ where $r$ is an integer
modulo $N$. The lattice is compactified, in the sense that the points $\bsx$ and
$\bsx + L$ are identified, $L$ being the lattice size:
\beq
L= a N.
\label{latticesize}
\eeq
A scalar field $\vph$ on the lattice is defined by a set of $N$ numbers
$\varphi_r$ ($r=1,2, \cdots , N$) which give the value on the field on the
lattice sites of coordinate $\bsx = r a$ :
\beq
\vph(\bsx) = \vph(r a) \equiv \varphi_r.
\label{scalarfield}
\eeq
The $N$ numbers $\vph_r$ can be regarded as the eigenvalues of a $N \times N$
diagonal matrix $\boldsymbol{\varphi}$:
\beq
\bsp = \begin{pmatrix}
\vph_1 & 0 & 0 & 0 & \cdots & 0 \\
0 & \vph_2 & 0 & 0 & \cdots & 0 \\
0 & 0 & \vph_3 & 0 & \cdots & 0 \\
\vdots & \vdots & \vdots & \vdots & \ddots & \vdots \\
0 & 0 & 0 & 0 & \cdots & \vph_N
\end{pmatrix} \label{matrixfield}
\eeq
whose rows and columns  are in one to one correspondence with
the sites of the lattice. Notice that the ordering of the rows and
columns is the same as the one of the lattice sites, so that
neighboring eigenvalues correspond to the values of the field in
neighboring sites. Derivatives are replaced on the lattice by
finite differences:
\beq
\frac{\partial \vph(x)}{\partial x} \rightarrow \frac{1}{a} \partial_{+}
\vph(\bsx) \equiv \frac{1}{a} \left( \vph(\bsx ) - \vph(\bsx -a ) \right)
\label{fdiff}
\eeq
or, with the notations of eq. (\ref{scalarfield}):
\beq (\partial_{+}\vph)_r = \left( \varphi_{r} -
\varphi_{r-1} \right) \label{finitediff}. \eeq
The breaking of the
translational invariance due to the discrete nature of the lattice
results into a violation of the Leibniz rule when the finite
difference of a product of two functions is considered. As
discussed in detail in ref. \cite{D'Adda:2004jb}, to which we
refer for a more exhaustive treatment, a modified Leibniz rule
holds in place of the usual one:
 \beq
 (\partial_{+}\varphi \psi)_r =
(\partial_{+}\varphi)_r \psi_r + \varphi_{r-1} (\partial_{+}\psi)_r.
\label{modleib1}
 \eeq
 In matrix notation finite differences may be represented using the
shift matrices $\bsD_+$ and $\bsD_-$:
\beq \bsD_+= \begin{pmatrix}
0 & 1 & 0 & 0 & \cdots & 0 \\
0 & 0 & 1 & 0 & \cdots & 0 \\
0 & 0 & 0 & 1 & \cdots & 0 \\
\vdots & \vdots & \vdots & \vdots & \ddots & \vdots \\
0 & 0 & 0 & 0 & \cdots & 1 \\
1 & 0 & 0 & 0 & \cdots & 0
\end{pmatrix}\;,
\qquad \bsD_-=\bsD_+^{-1}= \begin{pmatrix}
0 & 0 & 0 & \cdots & 0 & 1 \\
1 & 0 & 0 & \cdots & 0 & 0 \\
0 & 1 & 0 & \cdots & 0 & 0 \\
0 & 0 & 1 & \cdots & 0 & 0 \\
\vdots & \vdots & \vdots & \ddots & \vdots & \vdots \\
0 & 0 & 0 & \cdots & 1 & 0
\end{pmatrix}
\label{Delta}
\eeq
namely, in components
\beq
(\bsD_+)_{rs}=\d_{r,s-1}\;, \qquad (\bsD_-)_{rs}=\d_{r,s+1}\;.
\label{Deltacomp}
\eeq
In the continuum the derivative $\partial \varphi$ is just the
commutator $[\partial,\varphi]$; on the lattice however the
commutator $[ \bsD_+ , \bsp ]$ is not diagonal, its non vanishing
matrix elements being on a shifted diagonal as in $\bsD_+$. In
order to write the finite difference (\ref{finitediff}) as a
function defined on the lattice sites, namely a diagonal matrix,
we have to define it as: \beq (\partial_+ \bsp) = \bsD_- [ \bsD_+ ,
\bsp ]= \begin{pmatrix}
\vph_1-\vph_N & 0 & 0 & 0 & \cdots & 0 \\
0 & \vph_2-\vph_1 & 0 & 0 & \cdots & 0 \\
0 & 0 & \vph_3-\vph_2 & 0 & \cdots & 0 \\
\vdots & \vdots & \vdots & \vdots & \ddots & \vdots \\
0 & 0 & 0 & 0 & \cdots & \vph_N-\vph_{N-1}
\end{pmatrix}.\label{findiff} \eeq
The factor $\bsD_-$ in front of the commutator is responsible for
the violation of the Leibniz rule, in fact from the definition (\ref{findiff})
we have:
 \beq (\partial_+
\bsp \bspsi) = (\partial_+ \bsp ) \bspsi + \bsD_- \bsp \bsD_+ (\partial_+
\bspsi ), \label{modleib2} \eeq which is completely equivalent to
(\ref{modleib1}). Notice that $\bsD_- \bsp \bsD_+$ is a "shifted"
field, where the eigenvalue $\varphi_r$ has been replaced by
$\varphi_{r-1}$: $(\bsD_- \bsp \bsD_+)_r = \bsp_{r-1}$.
The correspondence between derivative and finite difference is not one-to-one.
In fact the finite difference can be defined as a right or a left difference,
 so that the following correspondence also holds:
 \beq
\frac{\partial \vph(x)}{\partial x} \rightarrow - \frac{1}{a} \partial_{-}
\vph(\bsx) \equiv - \frac{1}{a} \left( \vph(\bsx) - \vph(\bsx+ a) \right)
\label{fdiff2}
\eeq
The finite difference $(\partial_{-} \vph)$ is defined in terms of the shift
matrices as
\beq (\partial_- \bsp) = \bsD_+ [ \bsD_- ,
\bsp ], \label{findiff2} \eeq
and satisfies the modified Leibniz rule:
\beq (\partial_-
\bsp \bspsi) = (\partial_- \bsp ) \bspsi + \bsD_+ \bsp \bsD_- (\partial_-
\bspsi ),\eeq 
which shows that $\partial_-$ carries the same shift as $\partial_+$ but in the
opposite direction.
The modified Leibniz rule (\ref{modleib2}) reflects the fact that
translational symmetry on the lattice is a discrete, and not a
continuous symmetry. To make this point clear consider an action
given as the trace of a product of fields $\bsp_i$: \beq S = \Tr~~
\bsp_1 \bsp_2 \cdots \bsp_r. \label{traction} \eeq The trace
corresponds to the sum over all lattice sites, and translational
invariance can be simply expressed as the invariance of the trace under cyclic
permutations of the eigenvalues, namely
(\ref{traction}) under \beq \bsp_i \rightarrow \bsD_- \bsp_i
\bsD_+ = \bsp_i + \delta \bsp_i, \label{transl} \eeq where $\delta
\bsp_i  = - (\partial_+ \bsp)$ as defined in (\ref{findiff}). When the
r.h.s.\ of (\ref{transl}) is inserted into (\ref{traction}) all
orders of $\delta \bsp_i$ must be kept in order to preserve the
exact symmetry and the variation of the Lagrangian can be cast in
the form:
\begin{align}
\delta & \left( \bsp_1 \bsp_2 \cdots \bsp_r \right) = (\delta \bsp_1) \bsp_2
\cdots \bsp_{r-1} \bsp_r + (\bsp_1 + \delta \bsp_1) (\delta \bsp_2) \bsp_3 \cdots \bsp_r +
\nonumber \cdots \\ + &  (\bsp_1 + \delta \bsp_1)(\bsp_2 + \delta \bsp_2) \cdots
(\bsp_{r-1} + \delta \bsp_{r-1})(\delta \bsp_r), \label{modleib3}
\end{align}
which is again the modified Leibniz rule. It is clear that eq.~(\ref{modleib3})
follows directly from the invariance of (\ref{traction}) under
(\ref{transl}) by keeping all orders in $ \delta \bsp_i$ while
linear terms in $\delta \bsp_i$ give the ordinary Leibniz rule
typical of the continuum limit. We stressed this point because the
situation is different in the supersymmetric theory discussed in the following
sections: supersymmetry charges are non diagonal and hence supersymmetry
transformations of a product of superfields obey a modified
Leibniz rule on the lattice ~\cite{D'Adda:2004jb}, however these modified
Leibniz rules cannot be derived, at least in the present formulation, from a
field transformation as in (\ref{transl}).

In the approach of ref.~\cite{D'Adda:2004jb}, which we follow here, in order to
preserve all supersymmetries exactly on the lattice a shift is associated to
each (super)symmetry charge. Such shifts generate the extended lattice required
to describe supersymmetric theories  and each shift coincides with the one
appearing in the corresponding modified Leibniz rule.

\section{The $N=2$ one dimensional supersymmetric model on the lattice}

\subsection{The algebra}

The modified Leibniz rule discussed in the previous section for the non
supersymmetric case is a key  ingredient in the formulation of lattice
supersymmetry of ref.~\cite{D'Adda:2004jb}, which we will follow.
In this formulation a shift is associated to each (super)charge, and appears in
the corresponding modified Leibniz rule. So, if $\delta_{\alpha}$ is the
supersymmetry variation generated by the SUSY charge $Q_{\alpha}$ and
and $\Phi_i(\theta,x)$ ($i=1,2$) are superfields, the shift $a_{\alpha}$
associated to    $Q_{\alpha}$ will determine the modified Leibniz rule:
\beq
\delta_{\alpha}\left( \Phi_1(\theta,x)\Phi_2(\theta,x)\right) = \left(\delta_{\alpha}
\Phi_1(\theta,x)\right) \Phi_2(\theta,x) + \Phi_1(\theta,x+a_{\alpha}) \left(\delta_{\alpha}
\Phi_2(\theta,x)\right)
\label{mdsusy}
\eeq
The shifts generate the lattice on which the theory is defined, very much in the
same way as roots generate the root lattice. In the
supersymmetric case this lattice will have in general extra points and extra
links with respect to the bosonic case, the extra links corresponding to the
shifts of the supersymmetry charges. As in the case of the roots, non vanishing
(anti)commutators give rise to linear constraints among the shifts as required
by consistency with (\ref{mdsusy}). So, if for instance $\{ Q_{\alpha},
Q_{\beta} \} = P_{\mu}$ and we apply both sides of the equation to a superfield
product using the modified Leibniz rule (\ref{mdsusy}) we obtain
$a_{\alpha}+a_{\beta} = \pm n_{\mu}$, where $n_{\mu}$ is the shift corresponding
to a link in the $\mu$ direction and the sign ambiguity is related to the fact
that on the lattice $P_{\mu}$ may be represented by either $\bsD_+$ or $\bsD_-$
as discussed in the previous section.

The one dimensional supersymmetric model discussed in section \ref{the-model}
has two supersymmetry charges $Q_1$ and $Q_2$ satisfying the algebra
(\ref{susyalgebra}), notably $Q_1^2=Q_2^2= P_x$. If we denote by $a_{Q_1}$,
$a_{Q_2}$ and $a_{P_x}$ the shifts associated respectively to  $Q_1$, $Q_2$ and
$P_x$ we have:
\beq
2 a_{Q_1} = \pm a_{P_x},~~~~~~~~2 a_{Q_2} = \pm a_{P_x}.
\label{shifts}
\eeq
Since $2 |a_{Q_i}| = |a_{P_x}|$ it is convenient to use $|a_{Q_i}|$ as lattice
spacing $a=|a_{Q_i}|$ and associate to the derivative $\frac{\partial}{\partial
x}$ a finite difference of {\bf two} lattice spacings, namely in the matrix
notation:
\beq
\frac{\partial}{\partial x} \rightarrow \pm \frac{1}{2 a} \bsD_{\pm}^2,
\label{dspace}
\eeq
where the dimension $N$ of the matrix, namely the number of sites in the lattice
will then be assumed to be even.
As for the signs in eq.~(\ref{shifts}) there are essentially two possible
distinct choices, namely $a_{Q_1}$ and $a_{Q_2}$ equal or opposite.
Both choices are consistent from the point of view of the realization of SUSY
algebra on the lattice, we will choose  $a_{Q_1}=-a_{Q_2}$ which is the most
symmetric with respect to the reflection symmetry.
The non vanishing commutators of the algebra (\ref{susyalgebra}) are
replaced on the lattice by:
\beq
Q_1^2 = -\frac{1}{2 a} \bsD_-^2,~~~~~~~~~~Q_2^2 = \frac{1}{2 a} \bsD_+^2
\label{qsquare}
\eeq
and the representation of the supersymmetry charges in terms of the Grassmann
coordinates $\theta_i$ by:
\beq
Q_1 = \frac{\partial}{\partial \theta_1} - \frac{1}{2 a} \theta_1 \bsD_-^2
,~~~~~~~~~~Q_2=\frac{\partial}{\partial \theta_2} +
\frac{1}{2 a} \theta_2 \bsD_+^2.
\label{sch}
\eeq
Although formally written in the same way as in the continuum case, $\theta_i$
and $\frac{\partial}{\partial \theta_i}$ in eq.~(\ref{sch}) are not simply
anticommuting parameters and differential operators. In fact, for the
supersymmetry charges $Q_i$ in (\ref{sch}) to satisfy the modified Leibniz
rules (\ref{mdsusy}), it is necessary that $\theta_i$
and $\frac{\partial}{\partial \theta_i}$ themselves carry a shift.
This can be easily determined by requiring that all terms in each of eq.s
(\ref{sch}) carry the same shift. The result is that $\theta_1$ must contain a
shift factor $\bsD_+$ and $\theta_2$ a shift factor $\bsD_-$ (the opposite for
the $\frac{\partial}{\partial \theta_i}$). It is convenient then to represent
 $\theta_i$ as $4N \times 4N$ matrices acting on a space that is the direct
 product of the $N$ dimensional space of the lattice sites and of an internal
 $4$ dimensional space, which is in turn a direct product of two two-dimensional
 spaces in which $\theta_i$ are represented essentially by Pauli matrices:
\begin{align}
\t_1&\equiv L^{1/2}\s_+ \otimes \mathbf{1} \otimes \D_+\;,  & \t_2&\equiv
L^{1/2} \s_3 \otimes \s_+ \otimes \D_-\;, \label{ftheta0} \\
\frac{\prt}{\prt\t_1}& \equiv L^{-1/2}\s_- \otimes \mathbf{1} \otimes \D_-\;, &
\frac{\prt}{\prt\t_2}& \equiv L^{-1/2}\s_3 \otimes \s_- \otimes
\D_+\;;\label{ftheta}
\end{align}
where $L$ is the length of the lattice as defined in (\ref{latticesize}).

Eq.s\ (\ref{ftheta0}) and (\ref{ftheta}) can be written explicitly as:
\begin{align}
\t_1&\equiv L^{1/2}\begin{pmatrix}
0 & 0 & \D_+ & 0 \\
0 & 0 & 0 & \D_+ \\
0 & 0 & 0 & 0 \\
0 & 0 & 0 & 0
\end{pmatrix},
& \t_2&\equiv L^{1/2}\begin{pmatrix}
0 & \D_- & 0 & 0 \\
0 & 0 & 0 & 0 \\
0 & 0 & 0 & -\D_- \\
0 & 0 & 0 & 0
\end{pmatrix}, \label{theta}\\
& & & \non \frac{\prt}{\prt\t_1}&\equiv L^{-1/2}\begin{pmatrix}
0 & 0 & 0 & 0 \\
0 & 0 & 0 & 0 \\
\D_- & 0 & 0 & 0 \\
0 & \D_- & 0 & 0
\end{pmatrix},
& \frac{\prt}{\prt\t_2}&\equiv L^{-1/2}\begin{pmatrix}
0 & 0 & 0 & 0 \\
\D_+ & 0 & 0 & 0 \\
0 & 0 & 0 & 0 \\
0 & 0 & -\D_+ & 0
\end{pmatrix}, \label{dtheta}
\end{align}
where the entries of the above matrices are $N\times N$ matrices
and $\D_+$ and $\D_-$ are the shift matrices defined in (\ref{Delta}).

It is straightforward to check that the matrices (\ref{theta}) and
(\ref{dtheta}) satisfy the standard Grassmann algebra of the
$\theta$ variables. This matrix representation is quite general
and can be easily extended to an arbitrary number $n$ of variables
by using direct products of $n$ Pauli matrices, namely $2^n \times
2^n$ matrices.

The factors $L^{\pm 1/2}$ in (\ref{ftheta0}) and (\ref{ftheta}) have been introduced for dimensional
reasons. In fact it is clear from (\ref{qsquare}) and (\ref{sch}) that the
Grassmann variables $\t_i$ have dimensions of a square root of a length.
The size $L$ of the lattice is kept fixed in performing the continuum limit
$ a \rightarrow 0$ which  becomes equivalent to the ordinary large $N$ limit
in matrix models. All dimensional quantities can  be expressed in units of
$L$, which can then be consistently set to $1$. This notation will
be used, unless otherwise specified, in what follows. In these units the lattice
spacing $a$ is replaced by $\frac{1}{N}$ and the supercharges are given by:
\beq
Q_1 = \frac{\partial}{\partial \theta_1} - \frac{N}{2} \theta_1 \bsD_-^2
,~~~~~~~~~~Q_2=\frac{\partial}{\partial \theta_2} +
\frac{N}{2} \theta_2 \bsD_+^2,
\label{sch2}
\eeq
and the factors $L^{\pm 1/2}$ in the matrix expressions
(\ref{ftheta0}), (\ref{ftheta}), (\ref{theta}) and (\ref{dtheta}) for $\t_i$ and
$\frac{\prt}{\prt\t_i}$ are omitted.

\subsection{Fields and superfields}
The next ingredient we need in order to construct a supersymmetric lattice
theory is to introduce fields and superfields. Let us begin with the component
fields. As discussed in the previous
sections a scalar field is represented by a diagonal matrix in the $N$
dimensional lattice space. In order to provide its representation in the $4N$
dimensional space introduced above we need to use its (anti)commutation
properties with $\theta_i$ and $\frac{\partial}{\partial \theta_i}$ and hence
distinguish between bosonic and fermionic fields:
\begin{itemize}
\item Bosonic field: a field which commutes with all $\t$'s and
 $\frac{\prt}{\prt\t}$'s. A straightforward calculation gives
\begin{align} \label{phi}
\hat{\vph}\equiv \begin{pmatrix}
\vph & 0 & 0 & 0 \\
0 & \D_+\vph\D_- & 0 & 0 \\
0 & 0 & \D_-\vph\D_+ & 0 \\
0 & 0 & 0 & \vph
\end{pmatrix}
&&
\begin{aligned}
\vph& \equiv N\times N ~~ \text{matrix} \\
\t_i\hat{\vph}& =\hat{\vph}\t_i \\
\frac{\prt}{\prt\t_i}\hat{\vph}& =\hat{\vph}\frac{\prt}{\prt\t_i}
\end{aligned}
\end{align}

\item Fermionic field: a field which anticommutes with all $\t$'s
and $\frac{\prt}{\prt\t}$'s. A straightforward calculation gives
\begin{align} \label{psi}
\hat{\psiu}\equiv \begin{pmatrix}
\psiu & 0 & 0 & 0 \\
0 & -\D_+\psiu\D_- & 0 & 0 \\
0 & 0 & -\D_-\psiu\D_+ & 0 \\
0 & 0 & 0 & \psiu
\end{pmatrix}
&&
\begin{aligned}
\psiu& \equiv N\times N ~~ \text{fermionic matrix} \\
\t_i\hat{\psiu}& =-\hat{\psiu}\t_i \\
\frac{\prt}{\prt\t_i}\hat{\psiu}& =-\hat{\psiu}\frac{\prt}{\prt\t_i}
\end{aligned}
\end{align}
\end{itemize}

 Let us now introduce superfields. They can be defined as objects having the
 standard expansion in terms of $\t_i$, namely
 \beq
\Phi=\hat{\vph}+ i\t_1\hat{\psiu}_1+ i\t_2\hat{\psiu}_2+i \t_2\t_1\hat{D}\;,
\label{suexp} \eeq
or equivalently as a matrix commuting with all $\t$'s but not with
 $\frac{\prt}{\prt\t}$'s. Either way we get the following representation:
\beq
\Phi=\begin{pmatrix} \label{PHI}
\vph & -i\psiu_2\D_- & -i\psiu_1\D_+ & i D \\
0 & \D_+\vph\D_- & 0 &i \D_+\psiu_1 \\
0 & 0 & \D_-\vph\D_+ & -i\D_-\psiu_2 \\
0 & 0 & 0 & \vph
\end{pmatrix}.
\eeq
As discussed in the previous sections a scalar field is represented by a
diagonal matrix whose eigenvalues are the value of the field at the $N$ lattice
sites. A diagonal matrix has no shift attached to it, so a scalar field carries
no shift. We assume then that $\vph$ in eq.~(\ref{phi}) is diagonal.
In order for the superfield $\Phi$ to obey well defined modified Leibniz rules
we are forced to assume that all terms of the expansion at the r.h.s.\ of
(\ref{suexp}) be homogeneous to $\hat{\vph}$ and carry no shift.
This means that the fermionic fields $\psiu_1$ and $\psiu_2$ must carry shifts
opposite to the ones of $\t_1$ and $\t_2$, namely that they must be proportional
respectively to $\D_-$ and $\D_+$.
It is convenient then to introduce {\it diagonal} matrices $\psi_1$ and $\psi_2$
(not underlined Greek letters):
\beq
\psi_1 =\D_+ \psiu_1 , ~~~~~~~~~~~~~\psi_2 = \D_- \psiu_2,
\label{diagpsi}
\eeq
which satisfy the same relations (\ref{psi}) as the underlined ones.
In the matrix representation (\ref{PHI}) of the superfield all entries are then
diagonal matrices, and its explicit form now reads:
\beq
\Phi=\begin{pmatrix} \label{PHId}
\vph & -i\D_+ \psi_2 \D_- & -i \D_- \psi_1 \D_+ & i D \\
0 & \D_+\vph\D_- & 0 & i\psi_1 \\
0 & 0 & \D_-\vph\D_+ & -i\psi_2 \\
0 & 0 & 0 & \vph
\end{pmatrix}.
\eeq
This is the form of the superfield we will mostly refer to in what follows.

The last point to investigate is the continuum limit of the bosonic and
fermionic fields $\vph$ and $\psi_i$. We will assume that $\vph$ and any bosonic
field has a smooth continuum limit. More precisely we assume that the finite
difference $\vph_{r+1} - \vph_r$ is of order of the lattice spacing $a \sim
\frac{1}{N}$ in the continuum limit.
This means that the derivative is well defined even if the
finite difference is taken on a single lattice spacing. If this is true for any
bosonic field, it must be true also for the product of two fermionic fields, for
instance $\psi_1 \psi_2$: $(\psi_1)_{r+1} (\psi_2)_{r+1} - (\psi_1)_r (\psi_2)_r
= O(a)$ ($a \rightarrow 0$). On the other hand, since the continuum derivative
corresponds on the lattice to a finite difference over {\it two} lattice
spacings, for it to be well defined we only need to impose the smoothness
condition $(\psi_i)_{r+2} - (\psi_i)_r = O(a)$.
These two requirements leave us with two possibilities:
\begin{itemize}
\item A) $\psi_i$ are themselves smooth matrices, like the bosonic fields:
\beq
\D_- \psi_i \D_+ - \psi_i = O(a),~~~~~~~~~~a \rightarrow 0.
\label{smoothpsi}
\eeq
\item B) $\psi_i$ are smooth up to an alternating sign:
\beq
\D_- \psi_i \D_+ + \psi_i = O(a),~~~~~~~~~~a \rightarrow 0.
\label{smoothpsi2}
\eeq
\end{itemize}
In the latter B) case $\psi_i$ are not themselves smooth but can be written as
\beq
\psi_i = \mathbf{T} \psi_i^{(c)},~~~~~~~~~~~~{\rm (~case~~B~) },
\label{smth}
\eeq
where $\psi_i^{(c)}$ satisfies the condition (\ref{smoothpsi}), and $\mathbf{T}$
is the alternating $N \times N$ matrix:
\beq
\mathbf{T} = \begin{pmatrix}
1 & 0 & 0 & 0 & \cdots & 0 \\
0 & -1 & 0 & 0 & \cdots & 0 \\
0 & 0 & 1 & 0 & \cdots & 0 \\
\vdots & \vdots & \vdots & \vdots & \ddots & \vdots \\
0 & 0 & 0 & 0 & \cdots & -1
\end{pmatrix}. \label{Tmatrix}
\eeq
The matrix $\mathbf{T}$ anticommutes with $\D_{\pm}$, and hence with $\t_i$ and
$\frac{\partial}{\partial \t_i}$. By using (\ref{smth}) the matrix
representation for the (diagonal) fermionic field $\hat{\psi}$ takes the form:
\beq
\hat{\psi} = \mathbf{\hat{T}} \hat{\hat{\psi}},~~~~~~~~~{\rm (~case~~B~) },
\label{hatpsi}
\eeq
where
\beq
\hat{\hat{\psi}}= \begin{pmatrix}
\psi^{(c)} & 0 & 0 & 0 \\
0 & \D_+\psi^{(c)}\D_- & 0 & 0 \\
0 & 0 & \D_-\psi^{(c)}\D_+ & 0 \\
0 & 0 & 0 & \psi^{(c)}
\end{pmatrix}
\label{hathat}
\eeq
and
\beq
\mathbf{\hat{T}} = \begin{pmatrix}
\mathbf{T} & 0 & 0 & 0 \\
0 & \mathbf{T} & 0 & 0 \\
0 & 0 & \mathbf{T} & 0 \\
0 & 0 & 0 & \mathbf{T}
\label{tt}
\end{pmatrix}.
\eeq
Except for its entries being Grassmann odd $\hat{\hat{\psi}}$ has exactly the
same structure as the bosonic field $\hat{\vph}$ in (\ref{phi}) and hence it
commutes with $\t_i$ and $\frac{\partial}{\partial \t_i}$. The anticommuting
properties of $\hat{\psi}$ with respect to the $\t$s comes entirely from the
matrix $\mathbf{\hat{T}}$ in (\ref{hatpsi}).
Something similar happens in case A), where $\psi_i$ obeys the smoothness
condition (\ref{smoothpsi}) and hence, setting the convention that the
superscript $(c)$ always denotes a smooth matrix, we can write:
\beq
\psi = \psi^{(c)},~~~~~~~~~~~~~{\rm (~case~~A~) }.
\label{smth2}
\eeq
Eq.~(\ref{psi}) then takes the form, similar to (\ref{hatpsi})
\beq
\hat{\psi} = \mathbf{\hat{W}} \hat{\hat{\psi}},~~~~~~~~~{\rm (~case~~A~) },
\label{hatpsi2}
\eeq
where
\beq
\mathbf{\hat{W}}  = \left( \t_1 \frac{\partial}{\partial \t_1}-
\frac{\partial}{\partial \t_1} \t_1 \right)\left( \t_2 \frac{\partial}{\partial
\t_2}-
\frac{\partial}{\partial \t_2} \t_2 \right) = \begin{pmatrix}
\mathbf{1}&0&0&0\\0&-\mathbf{1}&0&0\\0&0&-\mathbf{1}&0\\0&0&0&\mathbf{1}
\end{pmatrix}. \label{ww}
\eeq

\subsection{Supercharges and susy transformations}

The two supercharges $Q_1$ and $Q_2$ were given on the lattice by eq.s
(\ref{sch2}). With the matrix representation for the Grassmann coordinates $\t_i$
introduced in the previous sections they can be written in the form:
\begin{equation} \label{superc}
\begin{aligned}
Q_1&=\frac{\prt}{\prt\t_1}- \t_1\frac{N \D_-^2}{2 }= \begin{pmatrix}
0 & 0 & -\frac{N \D_-}{2} & 0 \\
0 & 0 & 0 &-\frac{N \D_-}{2} \\
\D_- & 0 & 0 & 0 \\
0 & \D_- & 0 & 0
\end{pmatrix},\\
Q_2&=\frac{\prt}{\prt\t_2}+ \t_2\frac{N \D_+^2}{2}= \begin{pmatrix}
0 & \frac{N \D_+}{2} & 0 & 0 \\
\D_+ & 0& 0 & 0 \\
0 & 0 & 0 & -\frac{N \D_+}{2}\\
0 & 0 & -\D_+ & 0
\end{pmatrix}.
\end{aligned}
\end{equation}
The supersymmetry transformations of the superfield $\Phi$, defined in
(\ref{suexp}) and (\ref{PHId}), will be defined in the usual way as
\begin{align} \label{d1d2}
\d_1\Phi=\underline{\hat{\eta}}_1 [Q_1,~\Phi]\;, &&
\d_2\Phi=\underline{\hat{\eta}}_2 [Q_2,~\Phi]\;,
\end{align}
where $\underline{\hat{\eta}}_i$ are the superparameters of the
transformation that we are going presently to discuss.
First we notice that the supercharges $Q_1$ and $Q_2$ carry a shift as they
are proportional respectively to $\D_-$ and to $\D_+$. In order for
$\d_i\Phi$ to be shiftless like $\Phi$ itself it is necessary that the
superparameters $\underline{\hat{\eta}}_i$ also carry a shift opposite to
the one of the corresponding charges.
So $\underline{\hat{\eta}}_1$ must be proportional to $\D_+$ and
$\underline{\hat{\eta}}_2$ proportional to $\D_-$:
\beq
\underline{\hat{\eta}}_1 = \D_+ \hat{\eta}_1,~~~~~~~\underline{\hat{\eta}}_2
 = \D_- \hat{\eta}_2,
 \label{underleta}
 \eeq
 where $\hat{\eta}_i$ has no shift, namely all its entries are diagonal
 matrices. As in the case of the fermionic fields discussed in the previous
 section there are two different possibilities, that we labeled case A) and
 B), for  $\hat{\eta}_i$ to anticommute with the $\t_i$. We set:
 \beq
 \hat{\eta}_i = \mathbf{\hat{W}} \eta_i~~~~~~\rm{(case A)},~~~~~~~
 \hat{\eta}_i = \mathbf{\hat{T}} \eta_i~~~~~~\rm{(case B)},
 \label{etahat}
 \eeq
 where $\mathbf{\hat{W}}$ and $\mathbf{\hat{T}}$ are given respectively in
 eq.s (\ref{ww}), (\ref{Tmatrix}) and (\ref{tt}) and $\eta_i$ are just
 c-number Grassmann odd parameters.
 The supersymmetry transformations (\ref{d1d2}) are now completely
 determined. By expanding the superfield in $\t_i$ or equivalently by using
 the matrix representation (\ref{PHId}) one can write the supersymmetry
 transformations for each component. Again we have to distinguish the two
 possible discretizations, A) and B). In the case labeled A) we have for $\d_1$:
\begin{align}
\d_1\vph& =i \eta_1 \psi_1 &  \d_1D& =\eta_1 N\frac{\D_+^2[\D_-^2,~\psi_2]}{2 }
 \non \d_1\psi_1& =-i \eta_1 N\frac{\D_+^2[\D_-^2,~\vph]}{2} &
 \d_1\psi_2& =\eta_1 D, \label{delta1A}
\end{align}
and for $\d_2$ we have:
\begin{align}
\d_2\vph& =i \eta_2 \psi_2 &  \d_2D& =\eta_2 N\frac{\D_-^2[\D_+^2,~\psi_1]}{2}
 \non \d_2\psi_1& =-\eta_2 D & \d_2\psi_2& =i
 \eta_2 N\frac{\D_-^2[\D_+^2,~\vph]}{2}, \label{delta2A}
\end{align}
where $\psi_1$ and $\psi_2$ are the diagonal, smooth matrices defined in
(\ref{smth2})\footnote{We also remind the reader that all dimensional quantities
are expressed in units of the lattice length $L$, so that the lattice spacing
$a$ is just in these units $a=\frac{1}{N}$.}. The transformations (\ref{delta1A})
and (\ref{delta2A}) coincide with the ones given in (\ref{susycomp}) in the
continuum limit.
In the case labeled B) the supersymmetry transformations can be written in terms
of the smooth matrices $\psi_i^{(c)}$ defined in (\ref{smth}). The result is the
same as the one for the case A) except for a sign change in the susy
transformations of the bosonic fields:
\begin{align}
\d_1\vph& =-i \eta_1 \psi_1^{(c)} &  \d_1D& =-\eta_1 N \frac{\D_+^2[\D_-^2,
~\psi_2^{(c)}]}{2} \non \d_1\psi_1^{(c)}& =-i \eta_1 N \frac{\D_+^2[\D_-^2,~\vph]}{2} &
 \d_1\psi_2^{(c)}& =\eta_1 D, \label{delta1B}
\end{align}
and:
\begin{align}
\d_2\vph& =-i \eta_2 \psi_2^{(c)} &  \d_2D& =-\eta_2
N\frac{\D_-^2[\D_+^2,~\psi_1^{(c)}]}{2}
 \non \d_2\psi_1^{(c)}& =-\eta_2 D & \d_2\psi_2^{(c)}& =i
 \eta_2N\frac{\D_-^2[\D_+^2,~\vph]}{2} \label{delta2B}.
\end{align}

It can be easily verified that these susy transformations form a closed algebra. For instance if
\begin{align}
\d_2\Phi=\hat{\eta}_2 \D_-[Q_2,~\Phi]\;, &&
\d'_2\Phi=\hat{\eta}'_2\D_-[Q_2,~\Phi]\;,
\end{align}
we have \beq
(\d_2\d'_2-\d'_2\d_2)\Phi=2\eta'_2\eta_2\D^2_-\left\{Q_2,~[Q_2~\Phi]
\right\}=2\eta'_2\eta_2\D^2_-[\frac{N\hat{\D}^2_+}{2},~\Phi]
\eeq
where with a rather obvious notation $\hat{\D}_+$ is a $4 \times 4$ diagonal
block matrix whose diagonal entries are $\D_+$.
Likewise for $\d_1$:
\beq
(\d_1\d'_1-\d'_1\d_1)\Phi=2\eta'_1\eta_1\D^2_+[-\frac{N\hat{\D}^2_-}{2},~\Phi]\;,
\eeq
and \beq (\d_1\d_2-\d_2\d_1)\Phi=0. \eeq

The superparameters $\underline{\hat{\eta}}_i$ of the supersymmetry
transformations (\ref{d1d2}) carry a shift (\ref{underleta}), hence the susy
transformation of a product of superfields satisfies a
\emph{modified} Leibniz rule. For instance, if  we consider the variation
under $Q_1$ we have:
\beq \label{modleib}
\d_1(\Phi_1\Phi_2)=\hat{\eta}_1\D_+[Q_1,~\Phi_1\Phi_2]=(\d_1\Phi_1)
\Phi_2+(\D_+\Phi_1\D_-)\d_1\Phi_2\;,
\eeq
where $\D_+\Phi_1\D_-$ is a \emph{shifted superfield}, namely a superfield where
all eigenvalues of the diagonal matrices within $\Phi_1$ have been shifted of
one unit. A similar expression can be obtained for variations under
$Q_2$.
Supersymmetry transformations of products of an arbitrary number of superfields
are uniquely and unambiguously determined by eq. (\ref{modleib}).
However, when using modified Leibniz rules on the component transformations,
like for instance (\ref{delta1A}), some care has to be used.
In fact on the lattice, due to the shift carried by the $\t_i$,
\emph{superfields do  not commute}:
\beq
\Phi_1\Phi_2\ne\Phi_2\Phi_1. \label{nocomm}
\eeq
Nevertheless their first components, being diagonal matrices, do commute.
So, if we denote by $\Phi|_0\equiv\hat{\vph}$ the first
component of the superfield $\Phi$ in the $\theta$ expansion, namely its
 diagonal part in the matrix representation of eq.~(\ref{phi}), we have
\beq
(\Phi_1\Phi_2)|_0=(\Phi_2\Phi_1)|_0 \label{firstcomm}
\eeq
that is
\beq
\hat{\vph}^{(1)}\hat{\vph}^{(2)}=\hat{\vph}^{(2)}\hat{\vph}^{(1)}.
\label{firstcomm2}
\eeq
However if we compute susy transformations of $\vph^{(1)} \vph^{(2)}$ by using
the modified Leibniz rules and the component transformations (\ref{delta1A})
we find
\beq
\d_1 (\vph^{(1)} \vph^{(2)}) = \d_1\vph^{(1)}~~\vph^{(2)}+ \D_+ \vph^{(1)}\D_-
 \d_1\vph^{(2)} = i \eta_1 \left( \psi_1^{(1)} \vph^{(2)}+
  \D_+ \vph^{(1)}\D_- \psi_1^{(2)} \right), \label{fifi}
  \eeq
  whereas if we start from $\vph^{(2)} \vph^{(1)}$ we find:
\beq
\d_1 (\vph^{(2)} \vph^{(1)}) = \d_1\vph^{(2)}~~\vph^{(1)}+ \D_+ \vph^{(2)}\D_-
 \d_1\vph^{(1)} = i \eta_1 \left( \psi_1^{(2)} \vph^{(1)}+
   \D_+ \vph^{(2)}\D_- \psi_1^{(1)} \right). \label{fifi2}
  \eeq
  In spite of (\ref{firstcomm2}) the r.h.s.\ of (\ref{fifi}) and (\ref{fifi2})
  are different because the shifts act on different fields, although of course
  they would coincide in the continuum limit.
  In ref.~\cite{Bruckmann:2006ub} this was presented as an intrinsic ambiguity
  and inconsistency of the formulation, but it is not so.
  In fact $\hat{\vph}^{(1)}\hat{\vph}^{(2)}=\hat{\vph}^{(2)}\hat{\vph}^{(1)}$
  can be regarded as the first component of two distinct superfields
  $\Phi_1\Phi_2$ and $\Phi_2\Phi_1 $.  In the former case
  ($\hat{\vph}^{(1)}\hat{\vph}^{(2)}= (\Phi_1\Phi_2)|_0$) eq. (\ref{fifi})
  should be used, in the latter instead ($\hat{\vph}^{(1)}\hat{\vph}^{(2)}=
  (\Phi_2\Phi_1)|_0$) one should use eq.~(\ref{fifi2}). In any supersymmetric
  expression, like for instance the action, it is always possible to determine
  from the higher components to which superfield a particular product of
  component fields belongs to, so no ambiguity ever arises.
  Of course this problem can be avoided from the very beginning by using
  consistently the superfield formalism.

\section{The action}

We want now to construct an action on the lattice that reproduces eq.s
(\ref{action}) and (\ref{actioncomp}) in the continuum limit and is invariant
under the discrete susy transformations defined in (\ref{d1d2}).
Hereafter we concentrate on the case A) in (\ref{etahat}),
and the modification to the case B) is straightforward.
To begin with let us introduce the covariant derivatives $\cd_1$ and $\cd_2$,
defined as:
\begin{equation} \label{covder}
\begin{aligned}
\cd_1&=\frac{\prt}{\prt\t_1}+ \t_1 N\frac{\D_-^2}{2 }= \begin{pmatrix}
0 & 0 & N\frac{\D_-}{2} & 0 \\
0 & 0 & 0 &N\frac{\D_-}{2} \\
\D_- & 0 & 0 & 0 \\
0 & \D_- & 0 & 0
\end{pmatrix},\\
\cd_2&=\frac{\prt}{\prt\t_2}- \t_2 N\frac{\D_+^2}{2}= \begin{pmatrix}
0 & -N \frac{\D_+}{2} & 0 & 0 \\
\D_+ & 0& 0 & 0 \\
0 & 0 & 0 & N \frac{\D_+}{2}\\
0 & 0 & -\D_+ & 0
\end{pmatrix}.
\end{aligned}
\end{equation}
It is straightforward matrix algebra to check that these covariant
derivatives anticommute with each other and with all supercharges
\begin{align} \label{dq}
\{\cd_i,~Q_j\}=0\quad\forall~i,~j & & \text{and also} & &
\{\cd_1,~\cd_2\}=0.
\end{align}
Besides we have
\begin{align}
\cd_1^2=+N \frac{\hat{\D}^2_-}{2}\;, && \cd_2^2=-N \frac{\hat{\D}^2_+}{2}.
\end{align}
The superfield action (\ref{action}) in the continuum can be written on the
lattice in the same way, by simply replacing each symbol with the corresponding
matrices and the integral over $x$ with the trace:
\beq \label{latact}
S=\Tr\left(\left\{\frac{\prt}{\prt\t_1},~\left[\frac{\prt}{\prt\t_2},~\frac{1}{2}[\cd_2,~\Phi]
[\cd_1,~\Phi]+\mathrm{i}V(\Phi)\right]\right\}\right). \eeq
The invariance of $S$ under supersymmetry transformations on the lattice can
easily be proved. First we notice that $\frac{\prt}{\prt\t_1}$ and
$\frac{\prt}{\prt\t_2}$ in (\ref{latact}) can be replaced with respectively
$Q_1$ and $Q_2$ without affecting the trace, the extra terms being total
differences. So we can write:
\beq
S=\Tr\left(\left\{Q_1,~\left[Q_2,~\frac{1}{2}\Psi_2
\Psi_1+\mathrm{i}V(\Phi)\right]\right\}\right), \eeq
where we have defined the fermionic superfields
\begin{align}
\Psi_1=[\cd_1,~\Phi]\;, && \Psi_2=[\cd_2,~\Phi].
\end{align}
Consider now the variation $\d_1 S$ of the action defined according to
eq.~(\ref{d1d2}) and (\ref{underleta}):
\beq \label{d1s}\d_1 S =
\Tr\left(\left\{Q_1,~\left[Q_2,\hat{\eta}_1\D_+[Q_1,~\frac{1}{2}\Psi_2
\Psi_1+\mathrm{i}V(\Phi)]\right]\right\}\right). \eeq
By using Jacobi identities and eq.s (\ref{qsquare}) it is easily seen that the
expression under trace in (\ref{d1s}) is a total difference and hence it
vanishes when the trace is taken.
The invariance of the action is then proved in complete generality.

The lattice superfield action (\ref{latact}) can be expanded and written in
terms of component fields, by using the explicit matrix expression
(\ref{PHId}) of the superfield $\Phi$.
Let us consider first only the kinetic term, but including the mass term
(that is we set $g=0$ in the expression (\ref{potential}) ). The action
written in terms of component fields then reads:
\begin{align}
S = \Tr & \frac{1}{2} \left\{
-[N \frac{\D_+^2}{2},\vph][N\frac{\D_-^2}{2},\vph] - D^2 +
\D_-^2 [N \frac{\D_+^2}{2},\psi_1]\D_-\psi_1\D_+ + \D_+\psi_2\D_- \D_+^2
[N  \frac{\D_-^2}{2},\psi_2]\right\} \non
 -&\frac{m}{2}\Tr \left( \vph D+D \vph - i
\D_+\psi_2\D_-\psi_1 +i \D_-\psi_1\D_+\psi_2 \right).
\label{latact2}
\end{align}
In the continuum limit the action (\ref{latact2}) coincides with the one given
in (\ref{actioncomp}) with $g=0$.
The invariance of (\ref{latact2}) under susy transformations can be checked by
using  the component fields transformations (\ref{delta1A}) and (\ref{delta2A})
and the modified Leibniz rule. In agreement with the discussion at the end of
the last section, the order of the factors in each term of (\ref{latact2}) is
important in applying the modified Leibniz rule even if the factors
(anti)commute, and in fact the order in (\ref{latact2}) reflect the one that
each field had in the original superfield formulation  (\ref{latact}).
Also on the lattice the auxiliary field $D$ can be eliminated by a gaussian
integration leading to the action:
\begin{align}
S = \Tr & \frac{1}{2} \left\{
-[N\frac{\D_+^2}{2},\vph][N \frac{\D_-^2}{2},\vph] + m^2 \vph^2+
\D_-^2 [ N\frac{\D_+^2}{2},\psi_1]\D_-\psi_1\D_+ + \D_+\psi_2\D_- \D_+^2
[  N\frac{\D_-^2}{2},\psi_2]\right\} \non
 -&\frac{m}{2}\Tr \left( - i
\D_+\psi_2\D_-\psi_1 +i \D_-\psi_1\D_+\psi_2 \right),
\label{latact3}
\end{align}
which reduces to (\ref{actioncomp2}) (with $g=0$) in the continuum limit.

Let us introduce again the coordinates $\bsx = r a= \frac{r}{N}$
($r=1,2,\cdots,N$) on the lattice and represent the diagonal matrices in
(\ref{latact2})
by means on their eigenvalues, using the notation of eq.~(\ref{scalarfield}).
The trace becomes a sum over the coordinates
$\bsx$, and the action can be written as:
\begin{align}
S = \sum_{\bsx} & \frac{1}{2} \big\{ - \frac{ \vph(\bsx
+2a)\vph(\bsx)+ \vph(\bsx)\vph(\bsx+2a) - 2 \vph(\bsx) \vph(\bsx)}
{4 a^2}\non +& \frac{\psi_1(\bsx + a)\psi_1(\bsx) - \psi_1(\bsx)\psi_1(\bsx + a)
+\psi_2(\bsx + a)\psi_2(\bsx) - \psi_2(\bsx)\psi_2(\bsx + a)}{2a} -
D(\bsx)D(\bsx)  \non -m & \left[ \vph(\bsx)D(\bsx)+D(\bsx)\vph(\bsx) -
i \psi_2(\bsx +a)\psi_1(\bsx) + i \psi_1(\bsx)\psi_2(\bsx +a) \right]\big\}.
\label{latact4}
\end{align}
As in (\ref{latact2}), the order of the factors in each term has been preserved
and it reflects the original order of the superfield product. In this way the
action is invariant under supersymmetry transformations, using the modified
Leibniz rule whenever susy transformations are applied to products of fields.
In the coordinate notation of (\ref{latact4}) the supersymmetry transformations
(\ref{delta1A}) and (\ref{delta2A}) and the modified Leibniz rules are:
\begin{align}
\d_1\vph(\bsx)& =i \eta_1 \psi_1(\bsx) &  \d_1D(\bsx)& =-\eta_1
\frac{\psi_2(\bsx+2a)-\psi_2(\bsx)}{2 a}
 \non \d_1\psi_1(\bsx)& =i \eta_1\frac{\vph(\bsx+2a)-\vph(\bsx)}{2a} &
 \d_1\psi_2(\bsx)& =\eta_1 D(\bsx), \label{delta1Ax}
\end{align}
\begin{align}
\d_2\vph(\bsx)& =i \eta_2 \psi_2(\bsx) &  \d_2D(\bsx)&
=\eta_2\frac{\psi_1(\bsx)-\psi_1(\bsx - 2a)}{2a}
 \non \d_2\psi_1(\bsx)& =-\eta_2 D(\bsx) & \d_2\psi_2(\bsx)& =i
 \eta_2\frac{\vph(\bsx)-\vph(\bsx-2a)}{2a} \label{delta2Ax}
\end{align}
and
\begin{align}
\d_1\left(A(\bsx)B(\bsx)\right) &= \left(\d_1 A(\bsx) \right) B(\bsx) +
A(\bsx+a) \left(\d_1 B(\bsx) \right), \non
\d_2\left(A(\bsx)B(\bsx)\right) &= \left(\d_2 A(\bsx) \right) B(\bsx) +
A(\bsx-a) \left(\d_2 B(\bsx) \right). \label{susyx}
\end{align}
In writing the action and the supersymmetry transformations above we have used
the case A), namely we have chosen the fermion matrix fields $\psi_1$ and
$\psi_2$ to have a smooth continuum limit. In case B) a change of sign of the
fermionic fields on odd sites is required to obtain a smooth field and
performing the continuum limit. The result in the action (\ref{latact4}) would
be a sign change in front of the fermionic action which is not significant and
does not need to be discussed any further.
Correspondingly, as already reported in the previous section, a change of sign
would occur in the susy transformations of the fermionic fields in
(\ref{delta1Ax}) and (\ref{delta2Ax}).
The order of the different factors arising from the superfield expansion
corresponds to a symmetrization of the action, as it is apparent from
(\ref{latact4}). This is needed not just for the invariance of the action
under (\ref{delta1Ax}) and (\ref{delta2Ax}) with the modified Leibniz rule
(\ref{susyx}), but also for the Hamiltonian to be a
self-adjoint operator and hence the propagator to be real. Reflection positivity
is also explicitly satisfied by (\ref{latact4}).

The interaction term can also be calculated from the superfield expansion, and
in the same notation of (\ref{latact4}) is given by:
\begin{align}
S_{\it{int}} =& \frac{g}{4} \Big\{ -\vph^3(\bsx)D(\bsx)-
\vph^2(\bsx)D(\bsx)\vph(\bsx)-\vph(\bsx)D(\bsx)\vph^2(\bsx)-D(\bsx)\vph^3(\bsx)
\non +& i \vph^2(\bsx)\psi_2(\bsx+a)\psi_1(\bsx)+
i  \vph(\bsx)\psi_2(\bsx+a)\vph(\bsx+a)\psi_1(\bsx)+i
\psi_2(\bsx+a)\vph^2(\bsx+a)\psi_1(\bsx)\non+& i\psi_2(\bsx+a)\vph(\bsx+a)\psi_1(\bsx)\vph(\bsx)
+i \psi_2(\bsx+a)\psi_1(\bsx)\vph^2(\bsx) +i
\vph(\bsx)\psi_2(\bsx+a)\psi_1(\bsx)\vph(\bsx)\non-& i
\vph^2(\bsx)\psi_1(\bsx-a)\psi_2(\bsx)-
i  \vph(\bsx)\psi_1(\bsx-a)\vph(\bsx-a)\psi_2(\bsx)-i
\psi_1(\bsx-a)\vph^2(\bsx-a)\psi_2(\bsx)\non- &i\psi_1(\bsx-a)\vph(\bsx-a)\psi_2(\bsx)\vph(\bsx)
 -i \psi_1(\bsx-a)\psi_2(\bsx)\vph^2(\bsx)-i \vph(\bsx)\psi_1(\bsx-a)\psi_2(\bsx) \vph(\bsx) \Big\}.
\label{Sint}
\end{align}
The interaction term, written in the form (\ref{Sint}), is also invariant under
(\ref{delta1Ax}), (\ref{delta2Ax}) and (\ref{susyx}). All terms in (\ref{Sint})
can be obtained starting from just two terms, for instance
$ -\vph^3(\bsx)D(\bsx)$ and $i \vph^2(\bsx)\psi_2(\bsx+a)\psi_1(\bsx)$ by means
of a symmetrization procedure, that is by taking the factors in all possible
orders. However in exchanging two fields the arguments are shifted as if
different fields were carrying different shifts, namely $0$ for the bosonic
fields $\vph(\bsx)$ and $D(\bsx)$ and respectively $+a$ and $-a$ for the
fermionic field $\psi_2(\bsx)$ and $\psi_1(\bsx)$.
This symmetrization procedure can be used, in alternative to the superfield
expansion and with the same results, to generate the correct susy invariant
terms at the component level.

\section{Ward identities}
\subsection{Quadratic Ward identities}
The formalism developed in the previous sections has allowed us to construct a
one dimensional supersymmetric theory on the lattice with two supersymmetry
charges and  exact supersymmetry invariance under both of them.
Such invariance is in fact satisfied by any model constructed using  the
superfield (\ref{PHId}) and the covariant derivatives (\ref{covder}) as
ingredients.
Supersymmetry transformations on the lattice, just as translations, are discrete
and, when applied to a product of superfields, obey modified Leibniz rules
(\ref{modleib}). However, unlike translations, where the modified
Leibniz rule is equivalent to  keeping all orders of $\delta \vph_i$ as shown
in (\ref{transl}) and (\ref{modleib3}), supersymmetry transformations of a
product of superfields are not obtained by doing a substitution $\Phi \rightarrow \Phi
+ \delta_i \Phi$ on all terms of the product and keeping all orders of the
variation $\delta_i \Phi$.
This is an important point, as the derivation of Ward identities from a
symmetry of the action involves, in the functional integral formalism, a change
of variables under which the action and the integration volume are invariant.
Such change of variable does not exist for supersymmetry transformations on the
lattice, and the standard derivation of the Ward identities is not possible.
In this section and in the following one we shall investigate the possibility
that correlation functions obey {\it modified} Ward identities, defined in a way
that is reminiscent of the modified Leibniz rule. This is the case indeed for
the free theory (including mass term) to which this section is devoted, whereas
for the interacting theory the question is still open.

We begin with considering the quadratic action of eq.~(\ref{latact4}). In order
to avoid unnecessary complications we introduce a compact notation and write the
action as:
\beq
S = \frac{1}{2} \sum_{A,B} f_A M_{AB} f_B
\label{qact}
\eeq
where the indices $A$ and $B$ are composite indices, namely:
\beq
A \equiv (\alpha, x).
\label{cindex}
\eeq
The index $\alpha$ goes from $1$ to $4$ to denote the four types of component
fields $\{\vph,D,\psi_1,\psi_2\}$, while $x$ is just the $N$-valued space time
coordinate. We shall also define $|A|$ as two valued function that is $0$
(resp.\ $1$) if $f_A$ is a bosonic (resp.\ fermionic) field, so that we have:
\beq
f_A f_B = (-1)^{|A||B|} f_B f_A.
\label{bf}
\eeq
All the information about the action is contained in the matrix $M_{AB}$, which
in agreement with (\ref{bf}) can be chosen to satisfy the symmetry:
\beq
M_{AB} = (-1)^{|A||B|} M_{BA}.
\label{symM}
\eeq
Another symmetry property of $M_{AB}$ follows from the invariance of the action
under translations on the lattice. If we define an operator $T_+$ (and its
inverse $T_-$) that acts on the index $A$ shifting the coordinate $x$ of one
lattice spacing:
\beq
T_{\pm}A \equiv (\alpha, x \pm a)
\label{Tpm}
\eeq
then translational invariance implies:
\beq
M_{(T_+A)(T_+B)} = M_{AB}.
\label{TM}
\eeq
The supersymmetry transformations (\ref{delta1Ax}) and (\ref{delta2Ax}) can be
described in this notation by two matrices $\lambda^{(i)}_{AB}$ ($i=1,2$) and
read:
\beq
\delta_i f_A = \eta_i \lambda^{(i)}_{AB} f_B.
\label{susyc}
\eeq
The explicit expression of $\lambda^{(i)}_{AB}$ is not important here (it could
be easily derived from (\ref{delta1Ax}) and (\ref{delta2Ax})), the only formal
property we need is again translational invariance: $\lambda^{(i)}_{AB}=
\lambda^{(i)}_{(T_+A)(T_+B)}$; moreover it is worth mentioning that because of
its statistic changing nature $\lambda^{(i)}_{AB}$ is non vanishing only for
$|A|+|B|=1$.
The invariance of the action (\ref{qact}) under supersymmetry transformations
can then be translated into a set of conditions on the matrix $M_{AB}$. In fact
by applying the modified Leibniz rule we have:
\beq
\delta_i S = \sum_{AB} (\delta_i f_A) M_{AB} f_B + f_{(T_iA)} M_{AB}
(\delta_i f_B),
\label{vraction}
\eeq
where $T_i$ is $T_+$ for $i=1$ and $T_-$ for $i=2$.
By writing the variations explicitly by means of (\ref{susyc}), and using
symmetry properties of $M_{AB}$ and $\lambda^{(i)}_{AB}$ (including
translational invariance) one finds that (\ref{vraction}) is zero iff:
\beq
\left(M_{AB} + M_{(T_iA)B} \right) \lambda^{(i)}_{AC} + \left(M_{AC} +
M_{(T_iA)C} \right) \lambda^{(i)}_{AB} = 0,
\label{susycond}
\eeq
where the sum over $A$ is understood. One can check that (\ref{susycond}) is
satisfied by (\ref{latact4}), but it is completely general and depends only by
the assumption that the action is quadratic, translationally invariant and
invariant under supersymmetry transformations with the modified Leibniz rules.

In order to see how the condition (\ref{susycond}) can be
translated into a relation between correlation functions let us
introduce the generating functional of the correlation functions
$F(J)$: \beq F(J) = \int df_A e^{-\frac{1}{2} f_A M_{AB}
f_B + J_A f_A }, \label{genfun} \eeq where the integration
volume  $df_A$ is normalized so that $F(J=0)=1$. The generating
functional $F(J)$ can be calculated explicitly in this case, as
the defining integral is quadratic, and is given by: \beq F(J) =
e^{\frac{1}{2} J_A M^{-1}_{AB} J_B (-1)^{|A|} }. \label{genfun2} \eeq
Correlation functions are obtained by expanding $F(J)$ in powers
of $J$. In particular the only connected correlation function of a
free theory, the propagator, is given by: \beq < f_A f_B >
=\left. \frac{\partial}{\partial J_A} \frac{\partial}{\partial
J_B} F(J)\right|_{J=0} = M^{-1}_{AB}. \label{correl} \eeq
 We can now write (\ref{susycond}) as a relation among correlation
 functions by simply multiplying it by $M^{-1}_{BG} M^{-1}_{CF}$
 and use again translational invariance. This gives:
 \beq
 \lambda_{GA}M^{-1}_{AF} + \lambda_{GA}M^{-1}_{A(T_iF)}
 +\lambda_{FA}M^{-1}_{AG} +\lambda_{FA}M^{-1}_{A(T_i)G}=0.
 \label{susycond2}
 \eeq
 Replacing in (\ref{susycond2}) $M^{-1}_{AB}$ with the two points
 correlation function $<f_A f_B>$ and multiplying by the
 Grassmann odd parameter of the susy transformation $\eta_i$ we
 finally obtain:
 \beq
 <(\delta_i f_G) f_F+ f_{T_iG} (\delta_i f_F) +
 (\delta_i f_F) f_G+ f_{T_iF} (\delta_i f_G)> =
 < \delta_i \left( f_G f_F + f_F f_G \right)> =0,
 \label{ward}
 \eeq
 where the variation of the product in (\ref{ward}) is performed
 using the \textit{modified} Leibniz rule.

 It is interesting to compare the previous result with the one we
 would obtain by the standard functional integral method, namely
 by doing in the integral (\ref{genfun}) the substitution
 \beq
 f_A \rightarrow f_A + (\delta_i f_A)
 \label{subst}
 \eeq
 with $(\delta_i f_A)$ given by (\ref{susyc}). Such
 substitution however does not leave the action invariant because
 of the modified Leibniz rule in (\ref{vraction}), and one
 obtains:
 \beq
 < e^{J_A f_A} \left( J_A (\delta_i
 f_A)- \delta_{sub}S \right)> =0,
 \label{wardsub}
 \eeq
 where $\delta_{sub}S$ is the variation of the action under the
 substitution (\ref{subst}):
 \beq
\delta_{sub}S= \frac{1}{2} \left( f_B - f_{(T_iB)}\right)
M_{BA}(\delta_i f_A). \label{deltasub} \eeq
Notice that $\delta_{sub}S$, being proportional to $f_B -
f_{(T_iB)}$ is of order $\frac{1}{N}$ in the large $N$ continuum
limit. By expanding (\ref{wardsub}) in $J_A$ one obtains all
possible Ward identities. In particular, by considering the
coefficient of $J_A J_B$ of the expansion one gets: \beq <
(\delta_i f_A) f_B + f_A (\delta_i f_B) - f_A
f_B \delta_{sub}S > = 0. \label{wardsub2} \eeq
 The first two terms in (\ref{wardsub2}) would give the variation
 of $f_A f_B$ if the ordinary Leibniz rule were valid,
 while the last term is the direct consequence of the
 non-invariance of the action under the substitution
 (\ref{subst}). This term is quartic in $f_A$, and hence it
 gets contributions from disconnected correlation functions.
 It is then far from trivial that its effect would be to restore
 the symmetry by turning the ordinary Leibniz rule into the
 modified one of the "correct" Ward identities of eq.
 (\ref{ward}).

 The existence of modified Ward identities reflects a symmetry of the
 generating functional of the correlation functions $F(J)$. In fact the
 exponent at the r.h.s.\ of (\ref{genfun2}), namely the generator of the
 connected correlation functions, has the same structure as the quadratic
 action (but with $M^{-1}$ in place of $M$) and is invariant under similar
 transformations:
 \beq
 \delta_i J_C  = \eta_i (-1)^{|C|} J_F \lambda_{FC}
 \label{Jinv}
 \eeq
 provided a modified Leibniz rule (with $T_i$ replaced by $T_i^{-1}$) is
 applied when taking the variation of a product of sources $J_A$.
 Indeed it is easy to check that in this way the variation of the exponent in
 (\ref{genfun2}) under (\ref{Jinv}) is proportional to the l.h.s.\ of
 (\ref{susycond2}) which is vanishing thanks to the symmetry of the action.

\subsection{Momentum space}

In order to write explicitly the Ward identities associated to the quadratic
action (\ref{latact4}), it is convenient to go to the momentum space and define;
\beq
\tf_{\alpha}(p) = \frac{1}{N} \sum_{r=1}^N f_{\alpha}(r) e^{\frac{2i\pi}{N} pr},
\label{ftransform}
\eeq
where the integer $r$ and $\alpha$ define together the composite index\footnote{
We remind here that $\alpha=1,2,3,4$ labels the different types of component
fields: $f_{\alpha} \equiv \{\varphi,D,\psi_1,\psi_2\}$.}
$A\equiv \{\alpha, x=a r\}$ used in the previous subsection and
\beq
f_A \equiv f_{\alpha}(r).
\label{snot}
\eeq
The integer (modulo $N$) $p$ is the momentum measured in units of $\frac{1}{L}$.
The component fields in the momentum space will always be denoted, as in
(\ref{ftransform}), with the same letter as the original field surmounted by a
tilde.
Standard properties of momentum space follow directly from (\ref{ftransform}),
for instance the product of two local fields leads to the
well known convolution and momentum conservation:
\beq
f(x) = f_1(x) f_2(x) \rightarrow \tf(p) = \sum_{p_1,p_2} \tf_1(p_1) \tf_2(p_2)
\delta_{p,p_1+p_2}.
\label{conv}
\eeq
All the results of previous sections can be conveniently rephrased in this
language. The quadratic action (\ref{latact4}) for instance can be written as:
\begin{align}
S =&\sum_{p=1}^N \left\{ \frac{N^2}{2} \tp(p)\; \sin^2\frac{2\pi}{N}p \;\tp(-p) -
\frac{i N}{2} \tpsione(p)\; \sin\frac{2\pi}{N}p \;\tpsione(-p) \right. \non
 -&\frac{i N}{2} \tpsitwo(p)\; \sin\frac{2\pi}{N}p \;\tpsitwo(-p) - \frac{1}{2}
 \tD(p)\; \tD(-p) - \frac{m}{2} \left[ \vphantom{\frac{m}{2}}\tp(p)\; \tD(-p) +
 \tD(p)\; \tp(-p) \right. \non - i & \left. \vphantom{\frac{N^2}{2}} \left.
 \omega^{-p} \tpsitwo(p) \; \tpsione(-p) + i \omega^p
 \tpsione(p)\; \tpsitwo(-p)\vphantom{\frac{m}{2}} \right] \right\},
 \label{actmom}
 \end{align}
 where $\omega= e^{\frac{2i\pi}{N}}$.
 The action (\ref{actmom}) is invariant under the two sets of supersymmetry
 transformations; eq.s~(\ref{delta1A}) and (\ref{delta2A}),
    which in the momentum representation read:
 \begin{align}
\d_1\tp(p)& =i \eta_1 \tpsione(p) &  \d_1\tD(p)& =i \eta_1 N \omega^{-p}
\sin\frac{2\pi}{N}p \; \tpsitwo(p)
 \non \d_1\tpsione(p)& =  \eta_1 N \omega^{-p}\sin\frac{2\pi}{N}p \;\tp(p) &
 \d_1\tpsitwo(p)& =\eta_1 \tD(p), \label{delta1Am}
\end{align}
\begin{align}
\d_2\tp(p)& =i \eta_2 \tpsitwo(p) &  \d_2\tD(p)&
=-i \eta_2 N \omega^p \sin\frac{2\pi}{N}p \; \tpsione(p)
 \non \d_2\tpsione(p)& =-\eta_2 \tD(p) & \d_2\tpsitwo(p)& =
 \eta_2 N  \omega^p \sin\frac{2\pi}{N}p \; \tp(p). \label{delta2Am}
\end{align}
When acting on a product of fields the supersymmetry transformations must be
applied using the modified Leibniz\footnote{It is worth to remind that the
modified Leibniz rule is order sensitive, and that it can be applied to
component fields only when the order of the fields reflects the one of the
original superfields.}  rule (\ref{susyx}), which in the momentum
representation are:
\begin{align}
\d_1\left(\tf_{\alpha}(p) \tf_{\beta}(q) \right) &= \left(\d_1 \tf_{\alpha}(p)
 \right)\tf_{\beta}(q)  +
\omega^{-p} \tf_{\alpha}(p) \left(\d_1 \tf_{\beta}(q) \right) \non
\d_2\left(\tf_{\alpha}(p) \tf_{\beta}(q) \right) &= \left(\d_2 \tf_{\alpha}(p)
 \right)\tf_{\beta}(q)  +
\omega^{p} \tf_{\alpha}(p) \left(\d_2 \tf_{\beta}(q) \right).
 \label{susyxmom}
\end{align}
 The two points correlation functions can easily be obtained from the action
 (\ref{actmom}) and cast in a matrix form:
 \beq
 < \tf_{\alpha}(q) \tf_{\beta}(p) > = \delta_{q,-p} M^{-1}_{\alpha \beta}(p)
 \label{prop}
 \eeq
 with
 \beq
 \boldsymbol{M}^{-1}(p) =  \frac{1}{N^2 \sin^2\frac{2\pi p}{N}+ m^2}\begin{pmatrix}
1 & -m & 0 & 0 \\
-m & -N^2 \sin^2 \frac{2\pi p}{N} & 0 & 0 \\
0 & 0 & i N \sin \frac{2\pi p}{N} & -i m \omega^p  \\
0 & 0 & i m \omega^{-p} &  i N \sin \frac{2\pi p}{N}
\end{pmatrix}. \label{matrixprop}
\eeq
The argument of $\sin^2$ at the denominator in (\ref{matrixprop}) vanishes for
two values of $p$, namely $p=0$ and $p=\frac{N}{2}$. This means that there are
two poles in the Brillouin zone, namely that the doubling of the fermions is not
eliminated in this model but rather generalized to all types of fields to
preserve the balance between bosonic and fermionic degrees of freedom.
However one can see from (\ref{latact4}) that odd and even sites are coupled only
through the fermionic degrees of freedom, and are decoupled in the bosonic
lagrangian. So one set of bosonic degrees of freedom, say the ones on odd sites,
could be eliminated. This however would break exact supersymmetry, because
bosonic fields on odd sites would be obtained as a result of supersymmetry
transformation on odd sites fermions, as well as a result of the modified
Leibniz rule.

The supersymmetry transformations (\ref{delta1Am}) and (\ref{delta2Am}) can also
be written in a more compact notation by as
\beq
\delta_1 \tf_{\alpha}(p) = \eta_1 \lambda^{(1)}_{\alpha \beta}(p)
\tf_{\beta}(p),~~~~~~~~~~~~~~~~~~~~\delta_2 \tf_{\alpha}(p) = \eta_2
\lambda^{(2)}_{\alpha \beta}(p)\tf_{\beta}(p),
\label{delta12}
\eeq
where the $4 \times 4$ matrices $\boldsymbol{\lambda}^{(i)}(p)$ are given by:
\beq
    \begin{aligned}
 \boldsymbol{\lambda}^{1}(p) &=  \begin{pmatrix}
0 & \phantom{AA}0\phantom{AA} & i & 0 \\0 & 0  & 0 & i N
\omega^{-p}\sin \frac{2\pi p}{N} \\
N \omega^{-p}\sin \frac{2\pi p}{N} & 0 & 0 &0 \\
0 & 1 & 0& 0
\end{pmatrix},\\
\boldsymbol{\lambda}^{2}(p)&=\begin{pmatrix}
0 & 0 & 0 & i \\
0 & 0  & -i N \omega^{p}\sin \frac{2\pi p}{N} & 0 \\
0 & -1 & 0 &0 \\
N \omega^{p}\sin \frac{2\pi p}{N} & 0 & 0& 0
\end{pmatrix}.
\end{aligned}
\label{lambda12}
\eeq

The modified Ward identities (\ref{ward}) in the momentum representation
take the form:
\beq
\delta_i < \tf_{\alpha}(-p) \tf_{\beta}(p) + \tf_{\beta}(p) \tf_{\alpha}(-p)> =0.
\label{wardm}
\eeq
The variation in (\ref{wardm}) can be performed using eq.s (\ref{delta12}) and
the modified Leibniz rules (\ref{susyxmom}), leading to the equations:
\begin{align}
 \left( 1+\omega^{-p}\right) \lambda^{(1)}_{\alpha \gamma}(-p)< \tf_{\gamma}(-p)
 \tf_{\beta}(p)> + \left(1+\omega^p\right) \lambda^{(1)}_{\beta \gamma}(p)
 <\tf_{\gamma}(p) \tf_{\alpha}(-p)>=&0  \non
 \left(1+\omega^{p}\right)\lambda^{(2)}_{\alpha \gamma}(-p) <\tf_{\gamma}(-p)
 \tf_{\beta}(p)> + \left(1+\omega^{-p}\right) \lambda^{(2)}_{\beta \gamma}(p)
 <\tf_{\gamma}(p) \tf_{\alpha}(-p)>=&0.
 \label{wardm2}
\end{align}
These can be written explicitly in terms of the component fields, by assigning
specific values to the indices. From the first one, i.e. $\delta_1$ variation we
get:
\begin{align}
i <\tpsione(-p) \tpsione(p) > + N \sin \frac{2\pi p}{N} <\tp(-p) \tp(p)> &= 0
\non
-i N \sin \frac{2\pi p}{N}  <\tpsitwo(-p) \tpsitwo(p)> + <\tD(-p) \tD(p)>&=0
\non
i <\tpsione(-p) \tpsitwo(p) > + \omega^p <\tp(-p) \tD(p) > &=0.
\label{wardcomp1}
\end{align}
From the second equation in (\ref{wardm2}), namely the $\delta_2$ variation, we
get:
\begin{align}
i <\tpsitwo(-p) \tpsitwo(p) > + N \sin \frac{2\pi p}{N}  <\tp(-p) \tp(p)> &= 0
\non
i N \sin \frac{2\pi p}{N} <\tpsione(-p) \tpsione(p) > - <\tD(-p) \tD(p)>&=0
\non
i <\tpsitwo(-p) \tpsione(p)> - \omega^{-p} <\tp(-p) \tD(p) > &=0.
\label{wardcomp2}
\end{align}
Of course these identities can be checked directly using the explicit form of
the correlators, but they follow from the exact supersymmetry, endowed with the
modified Leibniz rule, discussed in the previous sections.

The next issue to be investigated will be if exact supersymmetric modified Ward
identities hold if the interaction is switched on and loop diagrams come into
the game.
The standard method based on performing the substitution (\ref{subst}) on the
functional integral in presence of sources would only lead on the lattice to
identities that contain, like (\ref{wardsub2}), the explicit variation of the
action under such substitution. In fact the action is supersymmetric invariant
only if the modified Leibniz rule is applied whenever the variation of a
product is taken, and the substitution (\ref{subst}) does not account for that.
The variation of the action under (\ref{subst}) is of order $\frac{1}{N} \equiv
a$ and is expected to vanish in the continuum limit, but the question is if this
contribution can be exactly accounted for by modifying the Ward identities as
shown above for the free theory.
This is a difficult task that will be left to future investigation.


\section{Conclusions}

In this paper we have shown in the simple one dimensional example
of $N=2$ supersymmetric quantum mechanics, that supersymmetry
transformations on the lattice can be defined without any
ambiguity with the aid of the modified Leibniz rule if the
superfield formalism is consistently used. This clarifies the
point raised in \cite{Bruckmann:2006ub} and overcomes their
objections. The other problem we approached in the paper is the
derivation of Ward identities, namely the problem of whether exact
supersymmetry is preserved at the quantum level. This is a non
trivial problem, because due to the modified Leibniz rule
supersymmetry transformations cannot be expressed as a change of
variables in the functional integral and the usual derivation of
the Ward identities would lead to extra terms\footnote{This
problem arises whenever modified Leibniz rule appear, as for
instance in field theories on non-commutative space-time
\cite{Dimitrijevic:2007cu,Aschieri:2008zv}.}, proportional to the
lattice spacing. We began to tackle this problem by showing that
at least in the case of the theory without interaction (but
including the mass term) exact "modified" Ward identities hold,
that reflect the modified Leibniz rule of the original symmetry.
This is of course a very preliminary step, as we have not been
able yet to apply the same procedure to the more interesting case
of the theory with interaction. So the problem is still open and
left to future investigation. We expect that the situation of
higher dimensional models will be similar. We already showed in
\cite{D'Adda:2005zk} that the $N=2$ supersymmetric Yang-Mills
theory in two dimensions can be formulated on the lattice in a way
that the action is exact with respect to the four nilpotent
supersymmetry charges, thus ensuring exact supersymmetry under all
of them. We also showed that $N=4$ supersymmetric Yang-Mills
theory in three dimensions can be formulated on the lattice in a
similar way \cite{D'Adda:2007ax}. However in all these cases the
exact symmetry is realized classically by using supersymmetry
transformations involving a modified Leibniz rule, and a
deformation of the Ward identities (if it exist!) will be needed
for an exact symmetry at the quantum level\footnote{Just before
completing this paper we received a paper by K. Nagata
\cite{Nagata:2008xk} where this problem is considered in the
context of a twisted two dimensional Wess Zumino model. }.

\section*{Acknowledgments}

We would like to thank I. Kanamori and K. Nagata for useful
discussions and comments.
This work is supported in part by Japanese Ministry of Education,
Science, Sports and Culture under the grant number 50169778 and
also by Istituto Nazionale di Fisica Nucleare (INFN) research funds.


\end{document}